\documentclass[twocolumn,english,floatfix,superscriptadress,prl,showpacs]{revtex4}

\usepackage{graphicx}
\usepackage{amsmath}
\usepackage{amssymb}
\usepackage{psfig}
\usepackage{bm}

\begin{document}

\title{Family of analytic entanglement monotones}

\author{A. Delgado}
\email{delgado@info.phys.unm.edu}
\affiliation{Department of Physics and Astronomy, University of New Mexico, 800
Yale Boulevard, 87131 Albuquerque, USA}
\author{ T. Tessier }
\email{tessiert@info.phys.unm.edu}
\affiliation{Department of Physics and Astronomy, University of New Mexico, 800
Yale Boulevard, 87131 Albuquerque, USA}

\begin{abstract}
We derive a family of entanglement monotones, one
member of which turns out to be the negativity. Two others are shown
to be lower bounds on the I-concurrence, and on the I-tangle, respectively [P. Rungta and C. M. Caves, to appear in Phys. Rev. A]. We compare these bounds with the I-concurrence and I-tangle on the isotropic states, and on rank-two density operators resulting from a Tavis-Cummings interaction. Our results provide a global structure relating several different entanglement measures. Additionally, they possess analytic forms which are easily evaluated in the most general cases.
\end{abstract}

\date{\today{}}


\maketitle

Entanglement has proven to be a useful resource in the implementation
of quantum information processing protocols \cite{Nielsen3}. This
observation, coupled with the important role played by entanglement
in the foundations of quantum theory, has led to the search for a
quantitative theory of entanglement. In this context, a deep connection
between the concepts of separability and of positive maps has been
established \cite{Peres,Horodecki}, and several different measures of entanglement have been
proposed \cite{MHorodecki}.

The concurrence and the tangle are two related entanglement measures
yielding analytic forms in the case of two qubit quantum systems \cite{Hill-Wootters}.
Due to their intimate connection with the entanglement of formation,
and with the phenomenon of entanglement-sharing \cite{Coffman}, they
have proven to be particularly useful tools for studying fundamental
issues in quantum mechanics. Recently, analytic forms for these entanglement
measures have been found for bipartite quantum systems of arbitrary
dimensions in an overall pure state \cite{Rungta1}. However,
the generalizations of these quantities, known respectively as the
I-concurrence and the I-tangle  \cite{Rungta2}, to mixed states of a bipartite system
of arbitrary dimensions, involves a difficult minimization over ensemble
decompositions. So far, it has been possible to calculate analytically
the I-concurrence and the I-tangle only for the isotropic states \cite{Rungta2},
and in the special case that the density matrix of an arbitrary bipartite
quantum system has a rank no greater than two \cite{Osborne}.

Except for the cases mentioned above, analytic forms for the I-concurrence
and the I-tangle are lacking. In this work we present a new family
of entanglement monotones, certain members of which constitute easily
computable lower bounds for the I-concurrence and for the I-tangle. In
order to construct these functions we make use of the fact that the
I-concurrence and the I-tangle are convex-roof extensions of the generalized
concurrence for pure states of arbitrary dimensions, and of the square
of this quantity, respectively. Therefore, the I-concurrence (I-tangle)
can be bounded from below by any convex function which agrees with
the I-concurrence (I-tangle) on the set of bipartite pure states.
This requirement is seen to be satisfied by specific members of the
set of entanglement measures introduced herein. Each is a simple function
of the negative eigenvalues generated via the partial transposition
operation, and may be easily calculated with any standard linear algebra
package. 

In order to estimate the quality of these functions as lower bounds,
we compare their values with the values of the I-concurrence and the
I-tangle on the family of isotropic states. The I-tangle, and the
corresponding lower bound, are also compared numerically for rank-two
density matrices arising in the context of the Tavis-Cummings model
\cite{Tavis}.

The construction that follows is based on the theory of majorization.
This formalism has become a very useful tool in characterizing the
relationships among density matrices, ensemble decompositions, and
measurement processes \cite{Schroedinger}, and has led to new insights in 
the structure of quantum algorithms \cite{Latorre}, and in the problem of 
Hamiltonian simulation \cite{Vidal1}.

The following is a brief review of the main tenets of majorization
theory. The reader is referred to \cite{Bhatia} for extensive
background on the subject. Given two vectors \( x \) and \( y \)
in \( {\mathbb R}^{n} \), we say that the vector \( x \) is majorized
by the vector \( y \), denoted by the expression \( x\prec y \),
when the following two conditions hold: \begin{eqnarray}
\sum _{i=1}^{k}x_{i}^{\downarrow } & \leq  & \sum _{i=1}^{k}y_{i}^{\downarrow },\forall k=1,\dots ,n\label{majorization_one} \\
\sum _{i=1}^{n}x_{i}^{\downarrow } & = & \sum _{i=1}^{n}y_{i}^{\downarrow }.\label{majorization_two} 
\end{eqnarray}
 Here, the symbol \( \downarrow  \) indicates that the vector coefficients
are arranged in decreasing order. 

Majorization is naturally connected
with the idea of comparative disorder \cite{Bhatia}. In fact, \( x\prec y \)
if and only if there exists a doubly stochastic matrix \( A \) such
that \( x=Ay \). A matrix \( A \) is doubly stochastic if its coefficients
\( a_{ij} \) are non-negative and \( \sum _{i}a_{ik}=\sum _{j}a_{kj}=1,\forall k \).
If we consider \( x \) and \( y \) to be probability distributions,
then the fact that \( x \) is majorized by \( y \) expresses the
idea that \( x \) is more disordered, in a quantifiable sense, than
\( y \).

In the case that only condition (\ref{majorization_one}) holds, we
say that \( x \) is weakly submajorized by \( y \). This is denoted
by the expression \( x\prec _{w}y \). We will make use of the following
two results concerning majorization \cite{Bhatia}: \begin{eqnarray}
x\prec _{w}y\in {\mathbb R}^{n} & \Rightarrow  & x^{+}\prec _{w}y^{+}\label{weak_relation_one} \\
x\prec _{w}y\in {\mathbb R}_{+}^{n} & \Rightarrow  & x^{p}\prec _{w}y^{p}, \forall p\geq 1,\label{weak_relation_two} 
\end{eqnarray}
 where the operations \( x^{p} \) and \( x^{+} \) act on each component
of \( x \) individually. The \( x^{+} \) operation simply converts
each of the negative entries in \( x \) into a zero.

The following relation allows us to construct a family of convex functions
of the negative eigenvalues of an hermitian matrix. Given any two
hermitian matrices \( A \) and \( B \), \begin{equation}
\label{hermitian_majorization}
\lambda (A+B)\prec \lambda ^{\downarrow }(A)+\lambda ^{\downarrow }(B),
\end{equation}
 where \( \lambda (T) \) denotes the vector whose coefficients are
eigenvalues of \( T \). Let us now define the vectors \( \tilde{\lambda }(T)=-\lambda (T)=\lambda (-T) \),
such that the negative coefficients in \( \lambda (T) \) become positive
in \( \tilde{\lambda }(T) \). Clearly Eq. (\ref{hermitian_majorization})
also holds for the vectors \( \tilde{\lambda }(T) \), \emph{i.e.,}
\begin{equation}
\label{tilde_majorization}
\tilde{\lambda }(A+B)\prec \tilde{\lambda }^{\downarrow }(A)+\tilde{\lambda }^{\downarrow }(B).
\end{equation}

Recognizing that the coefficients of \( \tilde{\lambda }(T) \) belong
to \( {\mathbb R}^{n} \), and applying property (\ref{weak_relation_one}),
Eq. (\ref{tilde_majorization}) becomes \begin{equation}
\label{tilde_weak_maj_plus}
\left( \tilde{\lambda }(A+B)\right) ^{+}\prec _{w}\left( \tilde{\lambda }^{\downarrow }(A)+\tilde{\lambda }^{\downarrow }(B)\right) ^{+}.
\end{equation}
 The coefficients of the vectors in Eq. (\ref{tilde_weak_maj_plus})
are, by definition, members of \( {\mathbb R}_{+}^{n} \). Thus, using
property (\ref{weak_relation_two}) we obtain \begin{equation}
\label{tilde_weak_maj_plus_P}
\left[ \left( \tilde{\lambda }(A+B)\right) ^{+}\right] ^{p}\prec _{w}\left[ \left( \tilde{\lambda }^{\downarrow }(A)+\tilde{\lambda }^{\downarrow }(B)\right) ^{+}\right] ^{p}.
\end{equation}
According to condition (\ref{majorization_one}) for \( k=n \), we have \begin{equation}
\label{tilde_ineq_k_n}
\sum _{i=1}^{n}\left( \tilde{\lambda }_{i}^{+}(A+B)\right) ^{p}\leq \sum _{i=1}^{n}\left[ \left( \tilde{\lambda }^{\downarrow }_{i}(A)+\tilde{\lambda }^{\downarrow }_{i}(B)\right) ^{+}\right] ^{p},
\end{equation}
 where we have removed the ordering of the vector on the left hand
side. The term inside square brackets on the right hand side of Eq.
(\ref{tilde_ineq_k_n}) can be bounded from above by \( \tilde{\lambda }_{i}^{\downarrow +}(A)+\tilde{\lambda }_{i}^{\downarrow +}(B) \),
yielding \begin{equation}
\label{Plus_bounded_ineq}
\sum _{i=1}^{n}\left( \tilde{\lambda }_{i}^{+}(A+B)\right) ^{p}\leq \sum _{i=1}^{n}\left( \tilde{\lambda }_{i}^{\downarrow +}(A)+\tilde{\lambda }_{i}^{\downarrow +}(B)\right) ^{p}.
\end{equation}
 Taking the \( p \)-th root of Eq. (\ref{Plus_bounded_ineq}), and using Minkowski's
inequality \cite{Bhatia}, we obtain \begin{eqnarray}
\left[ \sum _{i=1}^{n}\left( \tilde{\lambda }_{i}^{+}(A+B)\right) ^{p}\right] ^{1/p} & \leq  & \left[ \sum _{i=1}^{n}\left( \tilde{\lambda }_{i}^{+}(A)\right) ^{p}\right] ^{1/p}\nonumber \\
 & + & \left[ \sum _{i=1}^{n}\left( \tilde{\lambda }_{i}^{+}(B)\right) ^{p}\right] ^{1/p}.\label{Minkowsky_Ineq} 
\end{eqnarray}

The terms in square brackets on the right hand side of Eq. (\ref{Minkowsky_Ineq})
are the sums of the positive coefficients of \( \tilde{\lambda }(A) \)
\( ( \tilde{\lambda }(B))  \) to the \( p \)-th power,
or equivalently, to the sums of the absolute values of the negative
coefficients of \( \lambda (A) \) \( \left( \lambda (B)\right)  \)
to the \( p \)-th power. Thus, we see that the quantity \begin{equation}
\label{M_p_definition}
M_{p}(A)=\Big( \sum _{\lambda (A)<0}|\lambda (A)|^{p}\Big) ^{1/p}
\end{equation}
 obeys the triangle inequality on the set of hermitian matrices for
any \( p\geq 1 \). In particular, \( M_{p}(A) \) is a convex function,
\emph{i.e.,} \begin{equation}
\label{Convexity_of_M_p}
M_{p}(\alpha A+\beta B)\leq \alpha M_{p}(A)+\beta M_{p}(B)
\end{equation}
 for \( \alpha  \) and \( \beta  \) in the interval \( [0,1] \)
such that \( \alpha +\beta =1 \). Similar results may also be shown
to hold for the set of functions
\begin{equation}
\label{N_p_definition}
N_{p}\left( A\right) \equiv M_{p}\left( A\right) ^{p},\forall p\geq 1.
\end{equation}

The I-concurrence \( C(\rho ) \) of a density matrix \( \rho  \)
acting on a bipartite \( d \)-dimensional Hilbert space \( \mathcal{H} \)
is defined by \cite{Rungta2}\begin{equation}
\label{I-concurrence}
C(\rho )=\min _{\{p_{i},\Psi _{i}\}}\left\{ \sum _{i}p_{i}C(\Psi _{i})\Big |\rho =\sum _{i}p_{i}|\Psi _{i}\rangle \langle \Psi _{i}|\right\} ,
\end{equation}
 with the concurrence \( C(\Psi ) \) of a bipartite pure state \( |\Psi \rangle  \)
given by \begin{equation}
\label{ConcurrencePsi}
C(\Psi )=2\Big (\sum _{i<j}c_{i}^{2}c_{j}^{2}\Big )^{\frac{1}{2}}.
\end{equation}
 The \( c_{i} \)'s in Eq. (\ref{ConcurrencePsi}) are the coefficients
of the state \( |\Psi \rangle  \), written in the Schmidt decomposition
\cite{Nielsen3}. The I-concurrence is the convex-roof extension of
\( C(\Psi ) \), and represents the average value of the pure state
concurrences for an ensemble decomposition of \( \rho  \), minimized
over all possible ensemble decompositions. The I-tangle has a similar
construction, and is given by \begin{equation}
\label{I-tangle}
\tau (\rho )=\min _{\{p_{i},\Psi _{i}\}}\left\{ \sum _{i}p_{i}C^{2}(\Psi _{i})\Big |\rho=\sum _{i}p_{i}|\Psi _{i}\rangle \langle \Psi _{i}|\right\} .
\end{equation}

Due to a result by Uhlmann \cite{Uhlmann}, the I-concurrence
and the I-tangle are known to be the largest convex functions defined
on the set of density operators which agree with \( C(\Psi ) \) and
\( C^{2}(\Psi ) \), respectively, over the set of bipartite pure
states. Therefore, if we are able to find a convex function which
agrees with \( C(\Psi ) \) \( \left( C^{2}(\Psi )\right)  \) over
the set of bipartite pure states, then it will automatically constitute
a lower bound for the I-concurrence (I-tangle). This can be accomplished by observing that the coefficients in Eq.
(\ref{ConcurrencePsi}) are the absolute values of the negative eigenvalues
of the partial transpose of \( |\Psi \rangle  \). Since the partial
transposition operation is linear, the function \( M^{pt}_{2}(\rho ) \)
defined by \begin{equation}
\label{BICPT}
M^{pt}_{2}(\rho )=2M_{2}(\rho ^{pt})
\end{equation}
 is a convex function which agrees with the I-concurrence on the set
of bipartite pure states. Consequently, \( M^{pt}_{2}(\rho ) \) is
a lower bound for the I-concurrence, \emph{i.e.,} \begin{equation}
\label{CTPT}
M^{pt}_{2}(\rho )\leq C(\rho ).
\end{equation}
 In a similar way it can be shown that the function \( N^{pt}_{2}(\rho )\equiv [M^{pt}_{2}(\rho )]^{2} \)
is a lower bound for the I-tangle, \emph{i.e.,} \begin{equation}
\label{Tangle_lower_bound}
N^{pt}_{2}(\rho )\leq \tau (\rho ).
\end{equation}
 These results hold for arbitrary dimensional bipartite quantum systems.

A similar result was obtained in \cite{Verstraete} for the case of
two qubits where \( d=4 \). Specifically, it was shown that the negativity
\( {\mathcal{N}}(\rho ) \) is a lower bound on the concurrence \cite{Vidal2}.
The negativity is defined to be the absolute value of the sum of the
negative eigenvalues of the partial transpose of \( \rho  \). Thus,
the negativity is seen to be one member of our family of entanglement
monotones, \emph{i.e.,} \( {\mathcal{N}}(\rho )\equiv M_{1}^{pt}(\rho ) \).
Additionally, in the case \( d=4 \) the partial transpose of \( \rho  \)
has at most one negative eigenvalue, implying that \( M_{2}^{pt}(\rho ) \)
also reduces to the negativity in this situation.

The lower bounds given by Eqs. (\ref{CTPT}) and (\ref{Tangle_lower_bound})
are functions of the negative eigenvalues produced via the partial
transposition operation. Hence, they are entanglement monotones in
their own right. This follows directly from the monotonicity of \( {\mathcal{N}}(\rho ) \) \cite{Vidal2}. Specifically, it may be shown that monotonicity is preserved in Eqs. (\ref{M_p_definition}) and (\ref{N_p_definition}) for choices of \(p\) other than one.
Further, these quantities have the additional advantage that they may be evaluated
in a straightforward manner with the help of a standard linear algebra
package.

It has been shown that the positive partial transposition criterion
is a necessary and sufficient condition for separability for \( d\leq6  \)
\cite{Horodecki}. In higher dimensions, positivity under partial
transposition is a necessary, but not sufficient, condition for separability.
However, this is not a serious drawback for the usefulness of these lower
bounds. Indeed, it has been shown by numerical experiments and theoretical
results that the volume of the set of density operators with positive partial transpose
decreases exponentially with the dimension \( d \) of the
Hilbert space \cite{Zyczkowsky}.

The exact values of the I-tangle for the isotropic states \( \rho _{F} \)
was analytically calculated in \cite{Rungta2}. Isotropic states describe
a quantum system composed of two subsystems of equal dimension \( d \).
They are mixtures formed by the convex combination of a maximally
mixed state and a maximally entangled pure state, \emph{i.e.,} \begin{equation}
\label{Iso-state}
\rho _{F}=(1-\lambda )\frac{1}{d^{2}}I\otimes I+\lambda |\Psi ^{+}\rangle \langle \Psi ^{+}|.
\end{equation}
 Here, \( I \) is the identity operator acting on a \( d \)-dimensional
Hilbert space, and \( |\Psi ^{+}\rangle  \) is the state given by
\begin{equation}
\label{Psi_plus}
|\Psi ^{+}\rangle =\sum _{i=1}^{d}\frac{1}{\sqrt{d}}|i\rangle \otimes |i\rangle .
\end{equation}
 The parameter \( \lambda  \) in Eq. (\ref{Iso-state}) can be related
to the fidelity \( F \) of \( \rho _{F} \) with respect to the state
\( |\Psi ^{+}\rangle  \), (where \( F\equiv \langle \Psi ^{+}|\rho _{F}|\Psi ^{+}\rangle \in [0,1] \)),
via the relation \begin{equation}
\label{Lambda}
\lambda =\frac{d^{2}F-1}{d^{2}-1}.
\end{equation}
 It has been shown that the isotropic states are separable for \( F\leq 1/d \) \cite{MPHorodecki}.

\begin{figure}
{\centering \resizebox*{2.75591in}{!}{\includegraphics{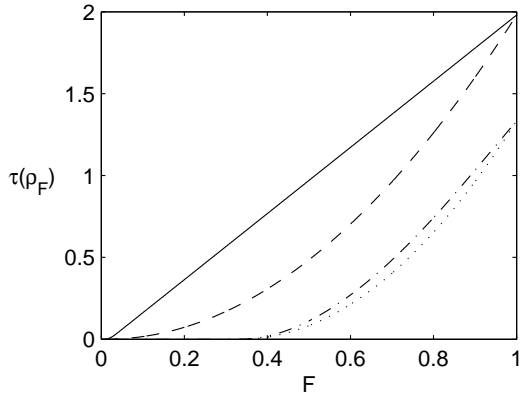}} \par}

\caption{Comparison of \protect\( \tau \left( \rho _{F}\right) \protect \)
and \protect\( N^{pt}_{2}\left( \rho _{F}\right) \protect \) as functions
of the fidelity \protect\( F\protect \) for different dimensions
\protect\( d\protect \). Solid line: \protect\( \tau \left( \rho _{F}\right) \protect \)
for \protect\( d=100\protect \). Dashed line: \protect\( N^{pt}_{2}\left( \rho _{F}\right) \protect \)
for \protect\( d=100\protect \). Dot-dashed line: \protect\( \tau \left( \rho _{F}\right) \protect \)
for \protect\( d=3\protect \). Dotted line: \protect\( N^{pt}_{2}\left( \rho _{F}\right) \protect \)
for \protect\( d=3\protect \).\label{ABB1}}
\end{figure}

The value of the lower bounds \( M^{pt}_{2}(\rho ) \) and \( N^{pt}_{2}(\rho ) \)
on the isotropic states can be calculated easily. Since the partial
transposition operation is linear, and since the identity operator
is invariant under this operation, the eigenvalues of \( \rho _{F}^{pt} \)
are readily obtained. They are given by \( (1-\lambda )/d^{2}\pm \lambda /d \)
with multiplicity \( d(d\pm1 )/2 \), respectively. The negative eigenvalues
\( (1-\lambda )/d^{2}-\lambda /d \) become positive when \( \lambda \leq1 /(d+1) \),
or equivalently, when \( F\leq 1/d \). Thus, \begin{equation}
\label{M_Isotropic}
M_{2}^{pt}(\rho _{F})=\left\{ \begin{array}{cc}
\frac{2}{d}\left( \frac{\lambda -1}{d}+\lambda \right) \sqrt{\frac{d(d-1)}{2}} & \lambda >1/(d+1)\\
0 & \lambda \leq1 /(d+1)
\end{array}\right. 
\end{equation}
 and \begin{equation}
\label{N_Isotropic}
N_{2}^{pt}(\rho _{F})=\left[ M_{2}^{pt}(\rho _{F})\right] ^{2}.
\end{equation}

The behaviors of the I-tangle \( \tau (\rho _{F}) \) and of \( N_{2}^{pt}(\rho _{F}) \)
for the isotropic states are depicted in Fig. \ref{ABB1}. For \( d=2 \),
the two functions assume the same values, while for larger dimensions
and constant fidelity, the difference between the lower bound and
the I-tangle increases. In the limit \( d\rightarrow \infty  \),
\( \tau (\rho _{F}) \) and \( N_{2}^{pt}(\rho _{F}) \) behave as
\( \sqrt{2}F \) and \( 2F^{2} \), respectively. Similarly, the I-concurrence
\( C(\rho _{F}) \) and \( M_{2}^{pt}(\rho _{F}) \) may be calculated
analytically. Here we find that the two quantities agree over the
isotropic states for any dimension \( d \), demonstrating that this
is a tight lower bound.

It has been shown \cite{Osborne} for the case of rank-two density
matrices that an analytic formula for the I-tangle exists. This class
of density matrices arises naturally in the context of the two-atom
Tavis-Cummings model (TCM). The TCM describes the interaction of an
ensemble of \( N \) two-level atoms with one mode of the quantized
electromagnetic field in the dipole and rotating wave approximations
\cite{Tavis}. In the case where \( N=2 \) and the initial state
of the overall system is pure, tracing over one of the atoms results
in a reduced density operator for the remaining atom and field subsystem
of rank at most two. Consequently, the entanglement between the field
and the atom can be quantified using Osborne's formula for the I-tangle \cite{Osborne}.
\begin{figure}
{\centering \resizebox*{2.75591in}{!}{\includegraphics{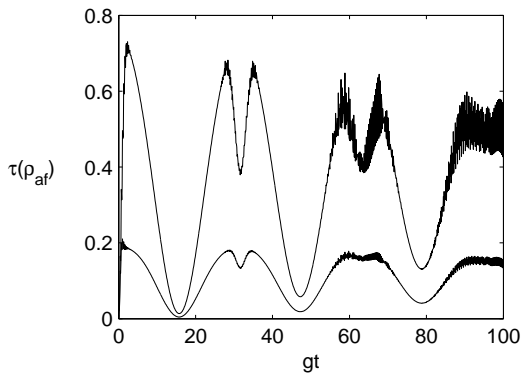}} \par}

\caption{Comparison of \protect\( \tau \left( \rho_{af} \right) \protect \) and
\protect\( N^{pt}_{2}\left( \rho_{af} \right) \protect \) calculated for
the two-atom TCM as functions of the effective time \protect\( gt\protect \).
The upper curve shows the evolution of \protect\( \tau \left( \rho_{af} \right) \protect \),
and the lower curve demonstrates the similar qualitative behavior
of the lower bound \protect\( N^{pt}_{2}\left( \rho_{af} \right) \protect \).\label{ABB2}}
\end{figure}

Fig. \ref{ABB2} shows the evolution of \( \tau \left( \rho _{af}\right)  \),
and of \( N_{2}^{pt}(\rho _{af}) \), as functions of the effective
time \( gt \), where \( g \) represents the coupling strength between
the two-level atoms and the electromagnetic field, and \( \rho _{af} \)
is the reduced density operator for one of the atoms and the field.
We have considered an ensemble of two atoms, both of which are initially
in the excited state. The field is initially in a coherent state with
an average photon number \( \left\langle n\right\rangle =100 \).
The I-tangle and its lower bound show clear differences in their magnitudes.
However, the lower bound preserves the basic structure present in
the evolution of the tangle. This agreement is a direct result of
the monotonicity of \( N_{2}^{pt}(\rho ). \)

The entanglement measures analyzed above belong to larger classes
of monotones defined by taking \( A=\rho ^{pt} \) in Eqs. (\ref{M_p_definition})
and (\ref{N_p_definition}). The negativity defined for a system of
two qubits is seen to be one member of this set. Other instances correspond
to lower bounds on the quantities I-concurrence and I-tangle, which
are useful tools for investigations of quantum information theoretic
concepts, and of fundamental quantum mechanics. Apart from offering
a larger structure from which to view these different entanglement
measures, this new family of functions also possesses analytic forms
which are easily computed even in the most general cases. It will
be interesting to see if further connections between seemingly unrelated
measures of entanglement can be found using this formalism.

The authors are grateful to I. Deutsch, A. Scott and P. Rungta for
many helpful discussions and comments. This work was supported under
the auspices of the Office of Naval Research, Grant No. N00014-00-1-0575.


\begin{thebibliography}{10}
\bibitem{Nielsen3} M. A . Nielsen and I. L. Chuang, {\it Quantum Computation and Quantum Information} (Cambridge University Press, Cambridge, England, 2000); G. Alber et al., {\it Quantum Information} (Springer, Berlin, 2001).
\bibitem{Peres}A. Peres, Phys. Rev. Lett. \textbf{77}, 1413 (1996).
\bibitem{Horodecki}M. Horodecki, P. Horodecki and R. Horodecki, Phys. Lett. A \textbf{223},
1 (1996).
\bibitem{MHorodecki} M. Horodecki, Quantum Inf. and Comput. \textbf{1}, 3
(2001).
\bibitem{Hill-Wootters}S. Hill and W. K. Wootters, Phys. Rev. Lett. \textbf{78}, 5022 (1997); W. K. Wootters, Phys. Rev. Lett. \textbf{80}, 2245 (1998).
\bibitem{Coffman}V. Coffman, J. Kundu, and W. K. Wootters, Phys. Rev. A. \textbf{61},
052306 (2000).
\bibitem{Rungta1}P. Rungta, V. Bu\v zek, C. M. Caves, H. Hillery and G. J. Milburn, Phys.
Rev. A \textbf{64}, 042315 (2001).
\bibitem{Rungta2}P. Rungta and C. M. Caves, to appear in Phys. Rev. A.
\bibitem{Osborne}T. J. Osborne, e-print quant-ph/0203087.
\bibitem{Tavis}M. Tavis and F. W. Cummings, Phys. Rev. \textbf{188}, 692 (1969). 
\bibitem{Schroedinger}E. Schr\"odinger, Proc. Cambridge Philos. Soc. \textbf{32}, 446 (1936); M. A. Nielsen, Phys. Rev. Lett. \textbf{83}, 436 (1999).
\bibitem{Latorre}J. I. Latorre and M. A. Mart\'{\i}n-Delgado, Phys. Rev. A \textbf{66},
022305-1 (2002).
\bibitem{Vidal1}G. Vidal and J. I. Cirac, Phys. Rev. A {\bf 66}, 022315 (2002). 
\bibitem{Bhatia}R. Bhatia, \textit{Matrix Analysis} (Springer, New York, 1997).
\bibitem{Uhlmann}A. Uhlmann, Phys. Rev. A \textbf{62}, 032307 (2000).
\bibitem{Verstraete}F. Verstraete, K. Audenaert, J. Dehaene and B. De Moor, J. Phys. A:
Math. Gen. \textbf{34}, 10327 (2001).
\bibitem{Vidal2}J. Eisert, Ph.D. thesis, University of Potsdam, (2001); G. Vidal and R. Werner, Phys. Rev. A \textbf{65}, 032314 (2002).
\bibitem{Zyczkowsky}K. \.Zyczkowski, P. Horodecki, A. Sanpera and M. Lewenstein, Phys. Rev.
A \textbf{58}, 883 (1998).
\bibitem{MPHorodecki}M. Horodecki and P. Horodecki, Phys. Rev.
A \textbf{59}, 4206 (1999).
\end{thebibliography}
\end{document}